\documentclass{svjour3}                   % Springer onecolumn (standard format)

\usepackage{array}
\newcolumntype{L}[1]{>{\raggedright\let\newline\\\arraybackslash\hspace{0pt}}m{#1}}
\newcolumntype{C}[1]{>{\centering\let\newline\\\arraybackslash\hspace{0pt}}m{#1}}
\newcolumntype{R}[1]{>{\raggedleft\let\newline\\\arraybackslash\hspace{0pt}}m{#1}}

\usepackage{url}
\usepackage{subcaption}
\usepackage{blindtext}
\usepackage{amssymb}
\usepackage{etoolbox}
\usepackage{booktabs}
\usepackage{graphicx}
\usepackage{graphics}
\usepackage{varioref}
\usepackage{placeins}
\usepackage{float}
\usepackage{xcolor}
\usepackage{multirow}
\usepackage{algorithmic}
\usepackage[ruled,vlined]{algorithm2e}
\usepackage[utf8]{inputenc}
\usepackage{hyperref}
\usepackage{caption}
\usepackage{float}
\usepackage{amsmath}
\usepackage{enumitem} 
\usepackage[numbers]{natbib}
\usepackage{adjustbox}

\def\tsc#1{\csdef{#1}{\textsc{\lowercase{#1}}\xspace}}
\tsc{WGM}
\tsc{QE}
\tsc{EP}
\tsc{PMS}
\tsc{BEC}
\tsc{DE}

\usepackage{etoolbox}
 
\newcommand*{\affmark}[1][*]{\textsuperscript{#1}}

\begin{document}
	%%%%%%%%%%%%%%%%%%%%%%%%%%%%%%%%%% Front Matter %%%%%%%%%%%%%%%%%%%%%%%%%%%%%%%
	\title
	{ Fast Griffin Lim based Waveform Generation Strategy for Text-to-Speech Synthesis
	}	
	\author
	{
		Ankit Sharma {\protect\affmark[a]} \and 
		Puneet Kumar {\protect\affmark[a]} \and 
		Vikas Maddukuri {\protect\affmark[b]\affmark[*]} \and 
		Nagasai Madamshetti\affmark[b] \and 
		Kishore KG\affmark[b] \and 
		Sahit Sai Sriram Kavuru\affmark[b] \and 
		Balasubramanian Raman \affmark[a] \and
		Partha Pratim Roy \affmark[a]
	}
	
	\authorrunning{Ankit Sharma et al.}

	\institute
	{
		\affmark[*] Corresponding author,  \email{mvikas@ec.iitr.ac.in} \\ 
		\affmark[a] Computer Science and Engg. Dept., Indian Institute of Technology, Roorkee, India, 247667 \\
		\affmark[b] Electronics and Comm. Engg. Dept., Indian Institute of Technology, Roorkee, India, 247667 
	}

	\date{Received: date / Accepted: date}
	
	\maketitle	
	\begin{abstract}
		The performance of text-to-speech (TTS) systems heavily depends on spectrogram to waveform generation, also known as the speech reconstruction phase. The time required for the same is known as synthesis delay. In this paper, an approach to reduce speech synthesis delay has been proposed. It aims to enhance the TTS systems for real-time applications such as digital assistants, mobile phones, embedded devices, etc. The proposed approach applies Fast Griffin Lim Algorithm (FGLA) instead Griffin Lim algorithm (GLA) as vocoder in the speech synthesis phase. GLA and FGLA are both iterative, but the convergence rate of FGLA is faster than GLA. The proposed approach is tested on LJSpeech, Blizzard and Tatoeba datasets and the results for FGLA are compared against GLA and neural Generative Adversarial Network (GAN) based vocoder. The performance is evaluated based on synthesis delay and speech quality. A 36.58\% reduction in speech synthesis delay has been observed. The quality of the output speech has improved, which is advocated by higher Mean opinion scores (MOS) and faster convergence with FGLA as opposed to GLA.
		\par
			
	\end{abstract}

	\maketitle
	
	%%%%%%%%%%%%%%%%%%%%%%%%%%%%%%%%% Introduction %%%%%%%%%%%%%%%%%%%%%%%%%%%%%%%%%%%	
	\section{Introduction}\label{sec1}
	The conclusive step in a text-to-speech (TTS) system is the generation of speech from the spectrogram representation of the signal. This process is known as the waveform reconstruction while the generation of intermediate signal from input text is called the construction process. The waveform generated in the reconstruction process is the time-domain signal obtained from its intermediate spectrogram. The overall performance of a TTS system depends on the waveform processing involved in the reconstruction phase \cite{mizuno1993waveform}. The main challenge in the TTS systems is to optimize the waveform processing time while maintaining or improving the quality of the generated speech \cite{zhao2018wasserstein}.\vspace{.01in}
	
	The TTS systems have emerged as valuable tools for day-to-day applications such as digital assistants, mobile phones, embedded devices, etc. Most of these devices have limited computational capacity and they are sometimes used in offline mode. Reducing the speech synthesis delay for them could be very useful in real-life applications such as - human computer interface, navigation systems, telecommunication and multimedia, aid to physically challenged people, daily appliances like TVs, washing machines, etc. \cite{jones2017real}. For such applications, TTS systems are expected to have quick response time and generate speech with good quality. Hence, it becomes even more important to fasten up spectrogram to speech reconstruction for real-time applications. With the aim to make TTS systems more suitable for real-time applications, it is important to improve their response time while retaining the quality of the synthesized speech.\vspace{.01in}
	
	Many software and hardware-based techniques have been suggested in the past for waveform optimization during speech synthesis \cite{r24, aaron2014systems}. Most of the traditionally used techniques were based on software components such as concatenative speech synthesis, parametric speech synthesis, etc. The current computing trend has shifted towards deep learning due to the availability of hardware resources and training data. The state-of-the-art of TTS systems have also started leveraging deep neural network (DNN) based techniques for speech synthesis \cite{wu2016google}. The performance of text-to-speech systems has significantly improved especially after the introduction of end-to-end neural waveform generation methods \cite{kim2017joint, jia2018transfer}. Inspite of significant performance boost, even end-to-end neural waveform generation approaches suffer from sluggish speech reconstruction process. Therefore, there is a need to look for more efficient waveform optimization approaches to enhance the speed and quality of machine synthesized speech.\vspace{.01in}
	 
	There are three stages in text-to-speech process: text analysis, linguistic analysis and waveform generation. Traditional TTS systems are based on complex multi-stage hand-engineered pipelines. The present state-of-the-art is end-to-end neural speech synthesis which puts together these stages of TTS process into a single layered pipeline through the use of DNNs. All three phases of TTS take place without human intervention for acoustic feature crafting. However, the hyper-parameters and configuration settings to cater a specific stage of the TTS can be set up-front. In that context, the implementation settings for waveform generation can be set in the form of appropriate choice of the reconstruction algorithm. Griffin-Lim algorithm (GLA) is the most predominantly used reconstruction algorithm for speech synthesis \cite{griffin1984signal}.\vspace{.01in}
	
	GLA is an iterative algorithm that tries to produce a signal from the spectrogram and does not have any information about phase. However, GLA needs many iterations and the perceptual quality of the output speech is not always very good \cite{masuyama2018griffin}. An optimized version of GLA is available in the literature which is known as Fast Griffin Lim Algorithm (FGLA) \cite{6701851}. FGLA naturally requires lesser iterations to construct the phase from spectrogram representation for general signal processing applications. However, it has not been applied and tested for speech synthesis. In this paper, we have applied FGLA for speech synthesis process of neural TTS systems. We've formulated an experiment to optimize the waveform processing of linear spectrograms in Tacotron TTS system. FGLA based reconstruction strategy has been applied to reduce the speech synthesis delay. And, observations have been made in context of the quality of the synthesized speech and the number of iterations required for the convergence of the reconstruction algorithm. \vspace{.01in}
	
	The proposed speech synthesis systems generate the speech from the magnitude spectral envelope. We have conducted a Mean opinion scores (MOS) study to test the quality of audio produced by FGLA with a lesser number of iterations and GLA with a greater number of iterations with an optimal number of training steps. The experiments have been conducted for LJSpeech, Blizzard and Tatoeba datasets. It resulted into 36.58\% reduction in speech synthesis time. The results have reflected higher quality of the output speech in terms of improved MOS. The number of training steps and iterations were determined by experimental observations. The convergence patterns of fourier transform plots of the resultant waveforms are found to be in-line with the choice of number of training iterations.
	
	\vspace{-.1in}
	\subsection{Contribution}\label{subsec1}
	The major contributions of the current research work are:
	\begin{itemize}		
		\item An FGLA based method has been proposed to reconstruct .wav speech files from linear spectrograms. In TTS applications, reconstruction of a waveform from spectrogram plays an important role because synthesis time is equivalent to the waiting time for application users. Users expect the speech output promptly. The proposed method has reflected into reduced synthesis time which is likely to enhance the experience of TTS application users.    
		\item The quality of the synthesis speech has been maintained while reducing the synthesis time. A market-based application cannot compromise about the quality of the synthesized speech. Speech quality also depends on the trained model. Hence, the model is trained upto optimal number of steps and the speech quality checking process has been carried out on three datasets. On all three datasets, FGLA based speech reconstruction produced better quality speech than GLA based construction.
		\item TTS models have been trained on LJSpeech, Tatoeba and Blizzard datasets and the waveform reconstruction has been carried out for GLA, FGLA and the GAN based vocoder. Optimal number of training steps and iterations have determined experimentally. The .wav files generated by the TTS models have been evaluated based on the quality of the output speech and the synthesis time. The speech quality has been analyzed by evaluating in terms of Mean Opinion Score (MOS) and the synthesis delay has been analyzed by measuring the time needed for the TTS model to synthesize the output speech.  
	\end{itemize}
	
	\vspace{-.1in}
	\subsection{Organization}\label{org}
	The rest of the paper is organized as follows. Existing work on waveform processing for TTS systems has been surveyed in Section \ref{lr}. Section~\ref{prob} formulates the problem statement. The details of the proposed methodology have been outlined in Section~\ref{mds}. Section~\ref{impl} presents the experimental setup. Analysis of the observed results has been presented in Section~\ref{rss}. Finally, Section~\ref{con} concludes the paper and highlights the scope for future research.

	%%%%%%%%%%%%%%%%%%%%%%%%%%%%%%%%% Literature Review %%%%%%%%%%%%%%%%%%%%%%%%%%%%%%%%%%%	
	\section{Literature Review}\label{lr}
	In recent years, text-to-speech processing has witnessed significant improvements. Traditionally, concatenative and parametric speech synthesis methods have been used for the task of text-to-speech conversion. In last couple of years, neural TTS systems have provided substantial performance boost in the quality of machine synthesized speech. A review of various research attempts in context of aforementioned methods alongwith the waveform optimization strategies followed by them have been provided in the following sections and their summary is presented in Table~\ref{tab:lr}.
	
	\subsection{Various Speech Synthesis Methods}
	\subsubsection{Concatenative Speech Synthesis}
	Concatenative models have dominated speech synthesis process since 1970's. They are based on searching and collecting small samples of speech components from the voice database \cite{541110}. The voice quality of the speech synthesized by them is more natural, however, they require a huge amount of voice database. There are two types of costs associated with them - a) searching cost and b) concatenative cost. Searching cost deals with searching specific voice segments corresponding to required broken portion and concatenative cost is related to joining these segments. As pointed by G. Coormanne et al. \cite{coorman2007speech} , one of the problems with these models is that they do not produce a good quality speech if a suitable match in the database corresponding to the required segment is not obtained. Another challenge with concatenative models is that they require complete dataset for generating a new set of voice. It is difficult to select the target unit from the voice database in order to minimize the difference between the required and selected samples \cite{tokuda2013speech}.
	
	\subsubsection{Parametric Speech Synthesis}
	Parametric speech synthesis is another widely used process to generate speech from text. TTS models use the same statistical models derived from the data \cite{zen2009statistical}. They follow a parameter generation approach unlike fetching the speech samples from the database. Hidden Markov Model (HMM) based TTS architectures are among the most famous parametric models. The primary step involved in them is to find a parametric form of speech including spectral and excitation parameters from the voice corpus and then model them by using a set of generative models \cite{tokuday2015directly}. In context of using HMM-based TTS systems, Y. Junichi et al. \cite{Yamagishi2009ThousandsOV} predicted the parameters then synthesized the speech for a given text. The benefit of this approach is that it does not require the complete dataset at synthesis time. T. Masuko et al. \cite{masuko1997voice} were success to change the speaker's voice easily using the parametric speech synthesis while L. Soojeong et al. \cite{lee2017spectral} tried statistical parametric method for enhancement of the speech. However, a disadvantage associated with parametric methods is that the voice quality of the synthesized speech is not as natural as in case of concatenative speech synthesis. 
	
	\subsubsection{Neural Speech Synthesis}	
 	Concatenative and parametric TTS systems have practical difficulties, for example, their different components need to be modeled and processed separately \cite{sotelo2017char2wav}. Deep learning-based end-to-end neural TTS systems, which are the current state-of-the-art solve this problem. They put together the intermediate stages of TTS process into a single, layered pipeline through the use of DNN which is carried out without human intervention for acoustic feature engineering. The recent neural TTS systems include Wavenet \cite{oord2016wavenet}, Char2Wav \cite{sotelo2017char2wav}, Tacotron \cite{wang2017tacotron}, Tacotron 2 \cite{shen2018natural}, DeepVoice \cite{arik2017deep}, DeepVoice 2 \cite{gibiansky2017deep}, DeepVoice 3 \cite{ping2017deep}, VoiceLoop \cite{taigman2017voiceloop}. Wavenet is based on a generative model that predicts samples based on probability distribution.	Tacotron \cite{wang2017tacotron} produces spectrograms from the text and then produces corresponding waveform using a vocoder. However, waveform generation has been a time-consuming process for the initial TTS systems and the speech output was not human-like. Use of improved spectrogram methods such as mel-spectrogram and better vocoders such as WORLD, GLA, etc. have helped solving these problems \cite{1164317}.\vspace{.01in}
	
	Char2Wav predicts the parameters of the WORLD vocoder and uses a SampleRNN conditioned upon WORLD parameters for waveform generation. WORLD \cite{morise2016world} on the other hand consists of three analysis algorithms for determining the fundamental frequency ($F_0$), spectral envelope and aperiodic parameters. Tacotron is another end-to-end model that uses seq-to-seq learning to map the text to spectrogram as intermediate data and then audio is generated by using vocoder. It uses Griffin Lim as the vocoder that generates audio waveforms from the linear spectrogram. It takes linear scale magnitude spectrogram and number of iterations as input and produces the corresponding waveform. It was observed that GLA in Tacotron converges in about 60 iterations \cite{wang2017tacotron}. Tacotron incorporates the GLA for phase estimation, followed by an inverse Short-Time Fourier transform (STFT) for waveform reconstruction. Tacotron 2 is an entirely neural network-based approach for speech synthesis which combines the seq-to-seq model feature used in Tacotron and generates the mel-spectrogram and performs speech synthesis using modified Wavenet vocoder.\vspace{.01in}
	
	In DeepVoice \cite{arik2017deep}, Wavenet architecture is modified and a fast synthesis system is developed during the audio synthesis stage. DeepVoice 2 \cite{gibiansky2017deep} is a multi-speaker model that has taken Tacotron architecture as a base and performed modification in Griffin Lim algorithm with Wavenet based vocoder. Deepvoice 3\cite{ping2017deep} is a fully convolutional attention-based neural end-to-end TTS system. Its architecture is capable of transforming several textual features into vocoder parameters such as mel-spectrograms, linear scale log spectrograms, spectral envelope, fundamental frequency ($F_0$), aperiodicity parameters, etc. These vocoder features are given as input to the waveform synthesis models. It uses three different vocoders - WORLD, Griffin-Lim and Wavenet. Both WORLD and Griffin Lim use linear spectrogram whereas the modified Wavenet in Tacotron2 uses mel-spectrogram for waveform synthesis. VoiceLoop \cite{taigman2017voiceloop} is an attention-based neural text to speech system referenced by a working memory model called phonological loop. It is capable of producing voices that are sampled in the wild. VoiceLoop replaces convolutional RNNs with memory buffer.\vspace{.01in} 
	
	In context of using neural vocodders, K. Oyamada et al. \cite{oyamada2018generative} focused on DNN based architecture to recover the phase information from magnitude spectrogram. K. Kumar et al. \cite{kumar2019melgan} proposed the MelGAN which is a fully convolutional, non-autoregressive vocoder. It generalized well for unseen speakers and showed significant speed up in speech construction from mel-spectrograms.	In a similar work, WaveGlow \cite{prenger2019waveglow} was proposed by replacing the vocoder part of Wavenet by deep neural architecture. It performed efficiently on large utterances but its performance degraded while converting small text samples into speech. TTS with neural vocoders such as WaveGlow have to repetitively go through serial steps of waveform construction which causes them to take more time while constructing small sentences. As observed by K. Oyamada et al. \cite{oyamada2018generative}, on CPU, some of the neural vocoders took three times longer than GLA for the speech synthesis. Their training time is as high as a few weeks and however their inference is fast on the GPU, the large size of the trained model makes their application very difficult with real-time devices having CPU with constrained memory. \cite{prenger2019waveglow, kumar2019melgan, oord2017parallel}. 
	
	\subsection{Waveform Processing in TTS systems}
	The final speech generated by the TTS systems is in the form of waveform while the intermediate representation is called spectrogram. The raw text input is converted into sampled embedding vector in the pre-processing phase, from which the intermediate frequency-time representation, that is, spectrogram is generated \cite{r06}. It is called the `Construction Phase'. Then, waveform is generated in the `Reconstruction Phase' using the vocoders. The efficiency of reconstruction algorithm majorly determines the overall performance of the TTS system. GLA has been the most predominantly used reconstruction algorithm for speech synthesis \cite{masuyama2018griffin}. A time-domain signal can be reconstructed from its amplitude spectrogram using the information about its phase. When no information is available about the phase and only the amplitude spectrogram is available, GLA is particularly suited for phase reconstruction. However, GLA needs many iterations and the perceptual quality of the output speech is not always very good \cite{arik2018fast}.\vspace{.01in}	
	
	There have been a number of attempts to optimize the waveform processing for speech reconstruction. For instance, Sercan et al. \cite{masuyama2018griffin} implemented transposed convolution layers alongwith non-linear interpolation which resulted into better utilization of modern multi-core processors than simple iterative strategy. In another work, Y. Fisher \cite{yu2015multi} used multi-scale context aggregation by dilated convolutions that resulted in simplified network alongwith increased state-of-the-art accuracy. In the context of waveform processing based applications, Z. Cheng and J. Shen et al. \cite{cheng2016effective} used the properties of the audio waveforms to recommend music based on the venue and surrounding of the user. As an attempt to enhance the vocoder module, M. Morise et al. \cite{morise2016world} proposed a new vocoder, WORLD for feature extraction and waveform synthesis. Y. Masuyama \cite{masuyama2019deep} proposed an enhanced phase reconstruction technique by combining DNN with GLA to build GLA-inspired neural network layers for waveform generation.\vspace{.01in}	
	
	Some of the distinctly related work in the area of signal processing maps to the utilization of fourier transformation techniques such as Gabor Transform \cite{bracewell1986fourier, malathi2017performance}. It is a special form of fourier transform that is used to determine the frequency and phase content of the signals represented in the form of spectrograms. In this direction, a real-time fast fourier transform algorithm was proposed by H. Sorensen et al. \cite{sorensen1987real}. Successful research has also been carried out to achieve phase recovery with lesser number of iterations as compared to GLA \cite{masuyama2018griffin}. It gives a hint to look out for alternative reconstruction algorithms requiring lesser iterations while maintaining the quality of the synthesized speech.\vspace{.01in}

	Though there have been various attempts to optimize the waveform processing in context of GLA. However, better alternatives for waveform reconstruction have not been explored to their best potential. An optimized version of GLA known as Fast Griffin Lim Algorithm (FGLA) is available in the literature \cite{6701851}. It requires lesser iterations to construct the phase from spectrogram representation for general signal processing applications. However, it has not been applied and tested for speech synthesis application. In this paper, FGLA based waveform generation method has been proposed with the aim to reduce synthesis delay. It aims to overcome the challenges faced by concatenative and parametric TTS systems by getting rid of the need of human intervention for acoustic feature engineering. Some of the challenges of using nerual vocoders for real-time TTS applications such as - slow speech synthesis with CPU, larger model size, complex architecture, etc. have also been considered and addressed.
	
	\begin{table}[]
		\centering
		\caption{Summary of literature review.}
		\begin{center}
			\resizebox{.9
				\textwidth}{!}{%
				\begin{adjustbox}{angle=90}
				\renewcommand{\arraystretch}{2} %for verticle spacing of the table
				\begin{tabular}{|p{2.3cm}|p{4cm}|p{3.2cm}|p{1.7cm}|p{2.2cm}|p{2cm}|p{2cm}|p{2.2cm}|}
					\hline
					\textbf{TTS Model}    & \textbf{Basic Method}  & \textbf{Properties} & \textbf{Text Preprocessing} & \textbf{Waveform Preprocessing} & \textbf{Spectrogram Type}   & \textbf{Construction} & \textbf{Reconstruction}   \\ \hline
					\textbf{Concatenative} & Searches the audio segment from speech database that is most relevant to the text & Simple to implement. Results into good quality speech      & -                  & -                      & -                  & -                       & -                \\ \hline
					\textbf{Parametric}    & Keeps track of parametric form of speech including spectral parameters            & Speaker voice can be changed with minimum efforts          & -                  & -                      & -                  & -                       & -                \\ \hline
					\textbf{Wavenet} \cite{oord2016wavenet}       & Uses dilated regressive CNN to predict present sample from past sample            & High quality. Can generate multi-speaker voice.            & Yes                  & No                      & Mel-spectrogram    & -                       & -                \\ \hline
					\textbf{Char2Wav} \cite{sotelo2017char2wav}      & Uses bidirectional RNN to produce waveform from textual content only              & Expert linguistic knowledge is not required                & No                  & Yes                      & Linear-spectrogram & -                       & SampleRNN        \\ \hline
					\textbf{Tacotron}\cite{oord2016wavenet}      & Encoder-decoder architecture based on RNNs                                        & Fully end-to-end; robust and fast processing               & Yes                  & Approximate            & Linear-spectrogram & CBHG                    & GLA              \\ \hline
					\textbf{Tacotron2}\cite{shen2018natural}    & Encoder-decoder architecture based on RNNs                                        & Better speech quality and smaller model size than Tacotron & Yes                  & Approximate            & Mel-spectrogram    & Convolution based       & Modified Wavenet \\ \hline
					\textbf{DeepVoice} \cite{arik2017deep}     & Uses five different DNNs for TTS; needs less parameters and faster than Wavenet   & Faster processing                                          & No                  & No                      & Linear-spectrogram & CBHG                    & GLA              \\ \hline
					\textbf{DeepVoice2} \cite{gibiansky2017deep}   & Wavenet based spectrogram to audio generation                                     & Can generate multi-speaker voices with less training       & No                  & Approximate            & Linear-spectrogram & Attention based encoder & GLA, Wavenet     \\ \hline
					\textbf{VoiceLoop} \cite{taigman2017voiceloop}     & Uses shifting buffer memory instead of RNNs                                       & Robust; produces lesser errors                             & Yes                  & Exact                  & Linear-spectrogram & Buffer shallow network  & WORLD            \\ \hline 
					\multicolumn{8}{l}{Here, GLA: Griffin Lim Algorithm; CBHG: (1-D convolution bank + highway network + bidirectional GRU)}\\
				\end{tabular}
			 
			\end{adjustbox}
			}
		\end{center}
		\label{tab:lr}%
	\end{table}%

	%%%%%%%%%%%%%%%%%%%%%%%%%%%%%%%%% Problem Formulation %%%%%%%%%%%%%%%%%%%%%%%%%%%%%%%%%%%	
	\section{Problem Formulation}\label{prob}
	The major objective of the proposed research work is to optimize the waveform generation process during speech synthesis by TTS systems. The speech synthesis time should be reduced without changing the quality of the output speech. The training phase for TTS device is performed once in a given system until there is a change in algorithm. The synthesis phase is executed on real-time speech synthesis devices having low computational power. This phase is repeated every time a text is converted into speech. System resources at training stage are generally of high computing power. However, most of the real-time speech synthesis systems have limited computing capabilities. Hence, the synthesis algorithm should take less amount of memory to make the speech synthesis more suitable for real-time applications. The problem statement is subjected to the following constraints:
	
	\noindent \textit{i)}	
	The average time taken ($T$) to convert the corresponding spectrogram to the waveform should be minimized. $T$ corresponds to the synthesis delay for n samples. \vspace{.05in}
	
	\noindent \textit{ii)}
	The number of iterations required ($itr$) for the output waveform to converge should be minimized. That is, their plots should reach to an optimal state as soon as possible. \vspace{.05in}
	
	\noindent \textit{iii)}
	Quality of the synthesized speech ($qual$) should be maintained. The reduction in speech synthesis time should not affect it. \vspace{.05in}
		
	\noindent \textit{iv)}
	The speech synthesis process should result into optimal resource utilization ($util$). It should cater to the limited computational resources of real-time TTS systems. \vspace{.1in}
	
	The aforementioned constraints can be modeled mathematically as shown in Eq.~\ref{eqb}. 
	
	{\fontsize{9}{8.75}\selectfont
		\begin{eqnarray}
		\label{eqb}
		\begin{split}		
		\centering
		&Subject\ to\ constraints:\vspace*{.4in}\\
		&\begin{cases} 
			minimize\ (T,\ itr)\\ 
			maximize\ (qual,\ util)\\ 
		\end{cases}\\\\
		&Where:\\
		&\begin{cases}
			n:number\ of\ inputs.\\
			itr:number\ of\ iterations\ required\ for\ the\ output\ waveform\ to\ converge.\\
			qual: quality\ of\ the\ synthesized\ speech.\\
			util: utilization\ of\ the\ computing\ capacity\ of\ TTS\ plateform.\\
			T = \{t_1 , t_2 , t_3 , ...... t_n\},\ time\ to\ convert\ coefficients\ into\ waveforms.\\
			t_1 , t_2 , ....... t_n\ are\ the\ times\ taken\ to\ convert\ the\ spectrograms\ to\ waveforms.\\ 						
			S: \{{u_1 , u_2 , u_3 ,...... u_n}\},\ set\ of\ spectrograms\ generated\ by\ the\ TTS\ system.\\
			u_1 , u_2 , ....... u_n\ are\ the\ coefficient\ matrices\ of\ the\ spectrogram.\\
			W = \{v_1,v_2 ,v_3,....... v_n\},\ set\ of\ corresponding\ waveforms\ generated\ by\ a\ vocoder.\\				
			v_1 , v_2 , ....... v_n\ are\ the\ waveforms\ produced\ from\ u_1 , u_2 , ....... u_n.\\
		\end{cases}		
		\end{split}
		\end{eqnarray}
	}   	
		
	%%%%%%%%%%%%%%%%%%%%%%%%%%%%%%%%% Methodology %%%%%%%%%%%%%%%%%%%%%%%%%%%%%%%%%
	\section{Methodology}\label{mds}
	In general, a TTS system contains three phases: a) text analysis (text to words), b) linguistic analysis (words to phonemes) and c) waveform generation (phonemes to sound). The first and second phases are carried out during the training phase. The TTS model is trained on text data and intermediate spectrogram is generated from the trained model for given input text. The third phase takes place during the synthesis when phase waveform is synthesized through this spectrogram. The proposed method aims to optimize the reconstruction of original speech signal from the intermediate spectrogram. Generally, a magnitude spectrogram doesn't contain the complete phase information. A reconstruction algorithm such as GLA iteratively recovers that information. GLA is an iterative algorithm that takes a high number of steps to recover the phase information. The proposed approach applies an optimized alternative FGLA, for phase reconstruction in waveform generation phase.\vspace{.01in}
	
\begin{figure}[]
		\begin{center}
			\includegraphics[width=1\textwidth]{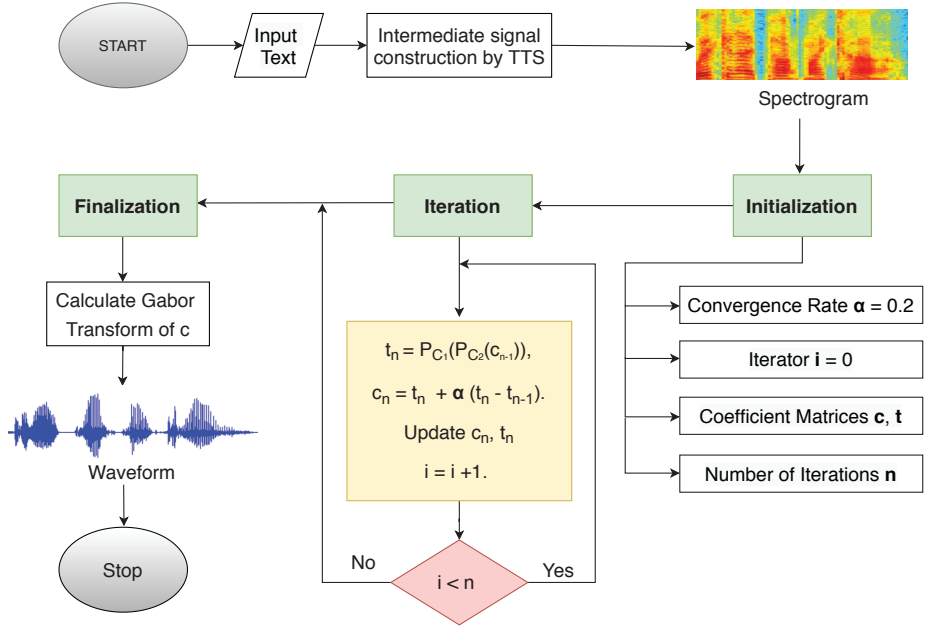}
			\caption{Flow diagram depicting the proposed methodology. Here, green colored boxes show various stages of the methodology; red box shows the loop decision and yellow box shows the loop iteration. Theoretical analysis of the complexity and the effect of convergence rate $\alpha$ is discussed in Section~\ref{strategy}. $\alpha$ is an important hyper parameter impacting the complexity and causing the speep-up in waveform processing. Its appropriate value is determined in Section~\ref{hparam}.}
			\label{flow}
		\end{center}
	\end{figure}

	The waveform analysis can be performed easily in the frequency domain \cite{bracewell1986fourier}. Fourier transform is most widely used transformation that converts time domain signal into frequency domain signal. That is why, the proposed methodology utilizes fourier transform and its variants such as Short-Time Fourier Transform (STFT), Discrete Fourier Transform (DFT), Gabor Transform, etc. during various steps of waveform generation.
	STFT is a series of fourier transforms of a subset of the signal. When frequency components of a signal vary with time, STFT is used to retrieve the time-localized frequency information. Gabor Transform is a special kind of STFT representation which is used to discover the phase information and sinusoidal frequency of the subsets of a time varying signal. The time-frequency analysis is carried out by first multiplying the function by a Gaussian function and then transforming it with a fourier transform. During the synthesis phase, input is a spectrogram and output is a waveform. The STFT can be represented as a matrix of coefficients where column index represents time and row index represents frequency of the respective DFT coefficient. The magnitude of each coefficient in the respective index is computed and this matrix can be treated as a image known as spectrogram of the signal. The choice of base implementation has been explained in Section~\ref{rationale}. The proposed methodology is described in the following sections. It is visually represented in Fig.~\ref{flow} and mathematically depicted in Algorithm~\ref{algo1}.
	
	\subsection {Rationale Behind Selecting Fast Griffin Lim Algorithm with Tacotron}\label{rationale}
	The research work presented in this paper primarily aims to reduce the speech synthesis delay in text-to-speech systems. A fundamental experiment to apply FGLA based waveform generation from linear-spectrogram has been formulated. The most commonly used waveform reconstruction algorithm for speech synthesis is GLA. The most acclaimed TTS system that uses linear-spectrogram for intermediate representation is Tacotron, developed by Google \cite{wang2017tacotron}. GLA was used as vocoder in Tacotron that produces approximate waveform corresponding to the input spectrogram, not the exact waveform. FGLA has been chosen instead of GLA because FGLA is known to take lesser iterations to recover the phase from spectrogram \cite{6701851}. However, other vocoders such as Wavenet  \cite{oord2016wavenet}, WORLD \cite{morise2016world}, etc. are available in the literature as potential choices for vocoders but they process mel-spectrograms while current research targeted to work with linear-spectrograms during the intermediate step of the speech synthesis process. Thus, FGLA with Tacotron emerged out as the most suitable choice for the experiment in consideration.

	\subsection{Strategy for Waveform Optimization}\label{strategy}
	STFT is used for the observable comparative analysis in the frequency domain \cite{griffin1984signal}. FGLA attempts to reconstruct the speech signal from the intermediate spectrogram of the signal. For that purpose, it finds the real signal $X^*\in R_L$ from a given set of spectral magnitude coefficients $s$, such that the magnitude of STFT of $X^\ast$ is as close as possible to the input signal. It helps in more accurate reconstruction of the signal. Any arbitrary set of complex numbers cannot be chosen as STFT coefficients, i.e., only a certain set of complex numbers correspond to STFT of a waveform. In the same way the input that we get may not be a valid spectogram. A valid spectrogram $S$ would have the magnitude of the coefficients as close as possible to the input.\vspace{.01in}
	
	The relevant terms have been defined in Section~\ref{prob}. Two important concepts that are utilized by the proposed method are Gabor Transform and Projection. $G^*x$ is the inverse Gabor Transform of $x$. It is a special case of STFT that is helpful in extracting the feature patterns from the spectrogram representation \cite{qian1993discrete}. It helps in finding the time needed to convert the spectrogram into waveform \cite{levoy1992volume}. Three phases of the proposed method have been described as follows.
		
	\subsubsection{Stage 1: Initialisation}
	\begin{enumerate}[label=(\roman*)]
		\item  First we initialize the coefficient matrix $c$, which is of same dimensions of the input spectrogram and contains the element of set $C_1$, $c_0$.

		\item The magnitude of every element is made equal to the element in input matrix in the corresponding position. That is, projecting on to set $C_2$. 
		
		\item Then we initialise another matrix $t$ of same dimensions. 
		
		\item The projection of the modified coefficients of the transform on to set $C_2$ is defined as follows.
		\begin{equation}
		P_{C_2}(c) = s.e^{i.\angle c}
		\end{equation}
		
		The above matrix is projected on to set $C_1$, followed by projection on to set $C_2$ to get $t_0$. Here, $t_i$ and $c_i$ denote the matrices after $i$ iterations.
		\begin{eqnarray}\fontsize{8}{6}\selectfont
			\begin{cases}
				C_0=s.e^{i.\angle({c_{0}})} \vspace{.1in}\\
				t_0=P_{C_2}(P_{C_1}(c_0))
			\end{cases}
		\end{eqnarray}

		\item Here, $C_1$ is the set of possible coefficients of STFT. $C_2$ is the set of complex numbers whose magnitude is equal to the magnitude spectrum coefficients. $G_x$ is the Gabor Transform of $x$.
		\begin{eqnarray}\fontsize{8}{6}\selectfont
			\begin{cases}
				C_1 = \{c :\exists  x \in R_L \|c=G_x\} \vspace{.1in}\\
				C_2 = \{c \in C_{MN} \| |c|=s \} 
			\end{cases}
		\end{eqnarray}

		\end{enumerate}
	
	\subsubsection{Stage 2: Iteration}	
	\begin{enumerate}[label=(\roman*)]
		\item The magnitude of the elements of the coefficient matrix is made equal to the input matrix, keeping the phase unchanged.
			\item Inverse Gabor Transform is then applied on the resultant coefficient followed by Gabor Transform. The projection of the modified coefficient on to the set $C_1$ is defined as follows. This is the influential step to make the FGLA faster than GLA \cite{6701851}.
		\begin{eqnarray}\fontsize{8}{6}\selectfont
			\begin{cases}
				P_{C_1}(c) = GG^*c \vspace{.1in}\\
				t_n=P_{C_1}(P_{C_2}(c_{n-1})) 
			\end{cases}
		\end{eqnarray}

		\item This projection on to set $C_1$ in current step is subtracted with the projection in set $C_1$ and multiplied by a factor, convergence rate $\alpha$. Choosing $\alpha$ close to one but not exactly one yields better results. This product is added to the projection in the current step, the initialised coefficient as updated to this value.		
		\begin{equation}
			c_n = t_n + \alpha(t_n - t_{n-1})
		\end{equation}
		
		\item The above steps are repeated iteratively and in each step the coefficients converge close to a real signal whose magnitude spectrum is approximately is equal to the input spectrum.\\
	\end{enumerate}

	\subsubsection{Stage 3: Final Waveform Generation}
	\begin{enumerate}[label=(\roman*)]
		\item Inverse Gabour Transform is applied on the final coefficient to get the waveform.		
		\begin{equation}
			x^* = G^*c_n
		\end{equation}
		
	\end{enumerate}\vspace{.25cm}

	The aforementioned phases of the proposed methodology are depicted in Algorithm~\ref{algo1} and the theoretical analysis of its complexity is discussed below.\vspace{.3cm} \\	
	\textbf{Complexity Analysis}\vspace{.2cm}\\
	As shown in Algorithm~\ref{algo1}, FGLA has single iteration loop and as per GLA paper \cite{griffin1984signal}, GLA also involves single loop. Hence, the theoretical time complexities for both FGLA and GLA are O(n). Experiments revealed that FGLA could produce the waveform of same quality with 30 iterations as compared to GLA with 60 iterations. Lesser number of iterations for FGLA is a determining factor for the reduction in the synthesis delay. Convergence rate $\alpha$ is another important hyper-parameter impacting the complexity and causing the speed-up in the waveform processing. Its value ranges from 0 to 1. For $\alpha$ = 0, FGLA behaves as GLA. As its value increases, speed-up also increases till a limit with faster convergence and then it starts decreasing. As discussed in Section~\ref{hparam}, the appropriate value of $\alpha$ has been determined as 0.2.

	\begin{algorithm}[]
		\caption{FGLA based Waveform Optimization for text-to-speech}
		\label{algo1}
		\textbf{Input} $n:$ Number of iterations.\\
		\textbf{Input} $k:$ Number of input variables.\\
		\textbf{Input} $C_1$: Set of possible coefficients of STFT. \\
		\textbf{Input} $C_2$: Set of complex numbers with magnitude same as of spectrum coefficients.\\
		\textbf{Define} $s:$ Spectral magnitude coefficients.\\
		\textbf{Define} $c:$ Coefficient matrix.\\
		\textbf{Define} $t:$ Matrix of same dimensions as c.\\
		\textbf{Define} $\alpha$: Convergence Rate.\\
		\textbf{Define} $x^*$: Final waveform.\\
		\textbf{Define} $G^*$: Inverse Gabor Transform of $x^*$.\\
		\textbf{Input} $S = \{{u_1 , u_2 , u_3, ..... u_k}\}:$ set of spectrograms generated by TTS model. \\
		\textbf{Define} $u_1 , u_2 , ....... u_k:$ coefficient matrices of the spectrogram. \\
		\textbf{Output} $W = \{v_1,v_2 ,v_3, ...... v_k\}:$ set of waveforms generated by a vocoder. \\
		\textbf{Define} $v_1 , v_2 , ....... v_k\:$: produced waveform.\\
		\textbf{Define} $T = \{t_1 , t_2 , t_3 , ...... t_k\}:$ time to convert coefficients into waveform. \\
		\textbf{Define} $t_1 , t_2 , ....... t_k\:$: time to convert the spectograms to waveforms.\\\vspace{.05in}
						
		\textbf{Procedure} WaveOpti
		\begin{algorithmic}[1] 
			\STATE \texttt{Stage 1: Initialisation}\\
			\STATE //\textit{Projection of modified transform coefficients on to set $C_2$}
			\STATE $P_{C_2}(c) = s.e^{i.\angle c}$\\
			\STATE // \textit{Projection of above matrix on set $C_1$ and then $C_2$} 
			\STATE $C_0=s.e^{i.\angle(c_0)}$\\
			\STATE $t_0=P_{C_2}(P_{C_1}(c_0))$.\\ \vspace{.08in}
			
			\STATE\texttt{Stage 2: Iteration}\\
			\STATE //\textit{Projection of modified transform coefficients on to set $C_2$}
			\STATE	$P_{C_1}(c) = GG^*c$\\
			\STATE //\textit{Update $t$ and $c$ for each iteration}
			\FOR{$i\ in\ n$}			
			\STATE	$t_i=P_{C_1}(P_{C_2}(c_{i-1}))$\\
			\STATE	$c_i = t_i + \alpha(t_i - t_{i-1})$.\\ 
			\ENDFOR			\vspace{.08in}

			\STATE\texttt{Stage 3: Waveform Generation}\\
			\STATE //\textit{Inverse Gabor Transform of final coefficients}
			\STATE	x$^*$ = G$^*$$c_n$
			
		\end{algorithmic}		
	\end{algorithm} \vspace{.2in}

	%%%%%%%%%%%%%%%%%%%%%%%%%%%%%%%%% Implementation %%%%%%%%%%%%%%%%%%%%%%%%%%%%%%%%%%%
	\section{Implementation and Results} \label{imps}
	This section discusses and evaluates the experiments to apply FGLA based waveform generation from linear-spectrogram.

	\subsection{Experimental Set-up}\label{impl}
	This section demonstrates the experimental implementation and analyses the results. A fundamental experiment to optimize the waveform processing of linear spectrograms in Tacotron TTS system has been formulated. The model training is done on Nvidia Tesla K80 GPU machine with 24GB RAM and 4992 CUDA cores. Text to speech synthesis is done on Intel(R) Core(TM) i7-7700, 4.2 GHz CPU with 16GB RAM and 64-bit Windows 10 OS machine. The Machine Learning libraries used in this implementation are Numpy, Tensorflow and Keras. The choice of parameters and datasets has been detailed in the following sections. 
	
	The questions that the experimental set-up tries to answer are - "What is the ideal number of training steps for TTS model training?"; "How to determine the appropriate number of iterations?"; "How to evaluate the speech synthesized by the TTS system in terms of quality, speed and convergence?" Suitable number of training steps and iterations are experimentally determined in Section~\ref{steps} and~\ref{itr}. Then as per Table~\ref{tab:tc}, use-case sentences are formulated according to various complexity levels. In Section~\ref{rss}, speech-synthesis for these sentences has been evaluated in terms of speech quality, synthesis delay and convergence. 
	
	\subsubsection{Hyper-parameter Selection}\label{hparam}	
	During the synthesis, the convergence rate $\alpha$ was experimentally chosen as 0.2. We experimented with the $\alpha$ values starting from 0.1 with learning rate 0.0002 and performed the iterations for spectrogram to waveform construction and analysed the construction time. The most suitable value of $\alpha$ corresponding to the optimal construction time emerged out as 0.2. The basic entity for training the model is .text and .wav file. Wav file signal has to be sampled for the analysis. Table~\ref{tab:pa} represents the important parameters for the current analysis and their selected values. Here, `Sampling rate' denotes the number of samples per second, `Frame shift' specifies amount by which window will slide. `Learning rate' shows how fast network learns by adjusting weights. The experimentally determined values for sampling rate, frameshift and learning rate are 20000, 12.5 ms and .0002 respectively. Tacotron has been trained from scratch for the datasets described in Section~\ref{dsets} for various number of iterations and speech synthesis time has been observed for GLA and FGLA both.
		
	\begin{table}[]
		\caption{Hyper-parameter choices.\vspace{-.2in}}
		\label{tab:pa}
		\begin{center}
			\renewcommand{\arraystretch}{1.25}
			\begin{tabular}{|l|c|c|c|}
				\hline
				\textbf{Parameter} & \textbf{Value} \\ \hline
				Convergence Rate, $\alpha$    & 0.2   \\ \hline
				Sampling Rate    & 20000   \\ \hline
				Frame Shift (ms) & 12.5    \\ \hline
				Learning Rate    & 0.0002  \\ \hline
			\end{tabular}
			
		\end{center}
	\end{table}

	\subsubsection{Datasets}\label{dsets}
	The original Tacotron paper had used LJSpeech dataset. We have trained and tested the TTS model with Blizzard and Tatoeba datasets as well. The details of all the datasets used in the implementation is provided in the following sections. 
	\begin{enumerate}[label=(\roman*)]
		\item  \textbf{LJ Speech Dataset} \cite{ljspeech17}: It is a single-speaker, public domain speech-dataset containing 13100 audio samples ranging from 1 to 10 seconds. The total duration of the dataset is about 24 hours. Each audio-file is a single-channel 16-bit PCM WAV with a sample rate of 22050 Hz. The properties of LJSpeech dataset are detailed in Table~\ref{tab:ljspeech}. \vspace{.05in} 
		
		\begin{table}
			\caption{Details of LJSpeech dataset \vspace{-.2in}}
			\label{tab:ljspeech}
			\begin{center}
				\renewcommand{\arraystretch}{1.25}
				
				\begin{tabular}{|p{4.2cm}|C{2.5cm}|}
					\hline
					\textbf{Parameters} & \textbf{Values}  \\ \hline
					Total Clips & 13,100\\ \hline
					Total Words & 225,715     \\ \hline
					Total Characters & 1,308,678   \\ \hline
					Total Duration  &	23:55:17 \\ \hline
					Mean Clip Duration &	6.57 sec \\ \hline
					Min Clip Duration &	1.11 sec \\ \hline
					Max Clip Duration &	10.10 sec \\ \hline
					Mean Words per Clip &	17.23 \\ \hline
					Distinct Words &	13,821 \\ \hline
				\end{tabular}
			\end{center}
		\end{table} 
	
	\item \textbf{Tatoeba Dataset} \cite{tatospeech}: This audio corpus is a crowdsourced dataset of sentences and translations. It contains a subset of the English sentences of Tatoeba. We have not used the complete dataset. Sentences have been filtered out and 299152 sentences have been included in processing. The average length of the audio clips ranges from 1 to 5 seconds. It is an open dataset. Various contributors keep on adding new audio clips and sentences.\vspace{.05in}
	
	\item  \textbf{Blizzard Dataset} \cite{Braunschweiler2010LightlySR}: It is available under Creative Commons Attribution Share-Alike license. It contains the samples of three audiobooks read by a single American English narrator. The books name and recording time are given in Table ~\ref{tab:blizzard}. Audio file format of blizzard corpus is 16-bit WAV, mono and sampling frequency is 44100 Hz.

		\linespread{1}
		\begin{table}[]
			\centering
			\caption{Details of Blizzard dataset\vspace{-.1in}}
			\label{tab:blizzard}
			\renewcommand{\arraystretch}{1.25}
			\begin{tabular}{|l|c|}
				\hline
				\multicolumn{1}{|l|}{\textbf{Audiobook Name}} & \multicolumn{1}{c|}{\textbf{Total Audio Length}} \\ \hline
				Tramp Abroad & 15:46:01 \\ \hline
				Life on the Mississippi & 14:47:27 \\ \hline
				\begin{tabular}[c]{@{}l@{}}The Man That Corrupted \\Hadleyburg and Other Stories\end{tabular} & 13:04:00 \\ \hline
			\end{tabular}%
		\end{table}
	
	\end{enumerate}

	\subsubsection{Determination of Appropriate Number of Training Steps}\label{steps}
	The quality of synthesized speech also depends on the number of training steps. However, training the model for more steps requires more computation and time. Hence, it is important to determine the optimal number of steps. We have trained the Tacotron model till 400K steps and observed the MOS values of the speech synthesized with it. MOS is a subjective evalution score to denote the quality of a speech utterance \cite{salza1996mos}. The MOS scores corresponding to various checkpoints are shown in Table~\ref{tab:ca}. It has been observed that the MOS values improves rapidly till 250k steps but their convergnce slows down significantly after that. Hence, the TTS model have been trained for at least 250k steps for the final training of each use-case. This analysis is performed with configuration same as original Tacotron paper, i.e., using GLA algorithm with 60 iterations.
	
	\begin{table}[]
	\caption{Determination of appropriate no. of steps.\vspace{-.2in}}
	\label{tab:ca}
	\begin{center}
		\renewcommand{\arraystretch}{1.25}
		\begin{tabular}{|c|c|c|c|}
			\hline
			\multicolumn{1}{|c|}{\multirow{2}{*}{\textbf{Steps}}}                                                                   
			& \multicolumn{3}{c|}{\textbf{MOS}}                                             \\  \cline{2-4} 
			&\textbf{LJSpeech} & \textbf{Tatoeba} & \textbf{Blizzard} \\\hline
			40k            & 7.0                     & 7.1                         & 5                      \\\hline
			80K            & 7.5                     & 7.3                         & 5.2                    \\\hline
			120K           & 7.6                     & 7.5                         & 5.4                     \\\hline
			160K           & 7.6                     & 7.6                         & 6.0                    \\\hline
			200K           & 7.65                    & 7.8                         & 6.3                    \\\hline
			240K           & 7.9                     & 8                           & 6.5                    \\\hline
			280K           & 8                       & 8.1                         & 6.7                    \\\hline
			320K           & 8.2                     & 8.25                        & 7                      \\\hline
		\end{tabular}
	\end{center}
	\end{table}
	
	\subsubsection{Determination of Appropriate Number of Iterations}\label{itr}	
	FGLA is supposed to converge faster than GLA. However, the suitable value for FGLA's number of iterations needs to be determined effectively. With that aim, we trained Tacotron with GLA and FGLA both for 20, 30 and 60 iterations respectively. Then we checked the MOS values for the audio samples synthesized with the model thus trained. These values have been illustrated in Table~\ref{tab:moe}. The speech quality for FGLA in terms of MOS scores is observed to be better than that for GLA. Moreover, FGLA is also observed to take lesser number of iterations to reach same to same MOS score as compared to GLA. FGLA with 30 iterations converged to equivalent MOS values for GLA with 60 iterations. Hence, 30 was selected as the appropriate number of iterations for FGLA to be used for Tacotron's training. The correctness of this choice has been justified in Section~\ref{res:time}.
	\begin{table}[]
		\caption{Determination of appropriate no. of iterations.\vspace{-.2in}}
		\label{tab:moe}
		\begin{center}
			\renewcommand{\arraystretch}{1.25}
			\begin{tabular}{|c|c|c|c|}
				\hline
				\textbf{Dataset}                     & \textbf{Iterations} & \textbf{GLA} 	& \textbf{FGLA}  \\\hline
				\multirow{3}{*}{\textbf{LJSpeech}}   & 20                  & 6.0            & 6.8            \\
				& 30                  & 7.3          & 8.2                 \\
				& 60                  & 7.6          & 8.2                 \\
				\hline
				\multirow{3}{*}{\textbf{Tatoeba}}   & 20                  & 7.6            & 7.1            \\
				& 30                  & 7.6          & 8.25         	   \\
				& 60                  & 8.1          & 8.25         	   \\
				\hline
				\multirow{3}{*}{\textbf{Blizzard}} 	 & 20                  & 6.0            & 6.2            \\
				& 30                  & 6.4          & 7.1            	   \\
				& 60                  & 6.9          & 7.3             	   \\\hline						
			\end{tabular}
		\end{center}
	\end{table}	
	
	\subsubsection{Use-case Formulation}
	During speech synthesis, the test sentences are chosen according to various complexity levels. Various verbal and lingual combinations in terms of punctuation marks, abbreviations, special characters, exclamation and question mark, etc. have been included to form five use-case sentences of varying lengths. As mentioned earlier, the main objective of the work presented in this paper is to reduce the synthesis delay without affecting the quality. The trained models are tested to synthesize these sentences and their synthesis delay speech quality has been observed. The time taken in the synthesis process is proportional to the length of the text. So, checked sentences have variable-sized length. Every sentence has been synthesized 10 times and then the average synthesis time has been considered. To make periodic observations, model checkpoint has been saved after every 1000 training steps. The size of the trained model has been observed to be of the order of 80 MB. The aforementioned use cases have been depicted in Table~\ref{tab:tc} and the detailed analysis has been presented in Table~\ref{tab:moe}.

	\linespread{1.25}
	\begin{table}[]
		\caption{List of use-case sentences.\vspace{-.2in}}
		\label{tab:tc}
		\begin{center}
			\renewcommand{\arraystretch}{1.25}
			\begin{tabular}{|p{.8cm}|p{6.5cm}|}
				\hline
				\textbf{S.No.} & \multicolumn{1}{c|}{\textbf{Sentences}} \\ \hline
				1.             & He said to him, “Is not your name Ahmed?" \\ \hline
				\multirow{2}[1]{*}{2.}            & All of a sudden, there was a loud screaming, Please help me! \\ \hline
				\multirow{2}[1]{*}{3.}            & I think I lost my wallet! I can’t find it anywhere! Oh, I could just kick myself! \\ \hline
				\multirow{2}[1]{*}{4.}            & "Sunshine on my shoulders makes me happy, sunshine in my eyes can make me cry." \\ \hline
				\multirow{2}[3]{*}{5.}            & As the stranger entered the town, he was met by a police, man who asked, “Are you a traveler?" “So it would appear", He replied carelessly. \\ \hline
			\end{tabular}
		\end{center}
	\end{table}

	%%%%%%%%%%%%%%%%%%%%%%%%%%%%%%%%% Results %%%%%%%%%%%%%%%%%%%%%%%%%%%%%%%%%%%
	\subsection{Result Analysis}\label{rss}
	The implementation has been carried out considering the number of training steps and iterations determined in the above section. This section presents and verifies the results in terms of speech synthesis using the TTS model thus trained. The results have been evaluated for five use-case sentences described in Table~\ref{tab:tc}. During the result evaluation, we have parallelly checked the quality of audio generated for all three corpora at different intervals of the model trained given in Table~\ref{tab:ca}. Mean opinion score (MOS) is calculated based on wave file generated by synthesis on the trained model. The optimal number of training steps for different datasets for various training steps are shown in the Table~\ref{tab:ca}. The results have been analysed based on the quality of the output speech and synthesis delay. The choice of number of iterations made in Section~\ref{impl} has also been verified by observing the convergence of the waveform plots for GLA and FGLA.
	
	\subsubsection{Quality Analysis}
\begin{figure}[]
		\begin{center}
			\includegraphics[width=0.67\textwidth]{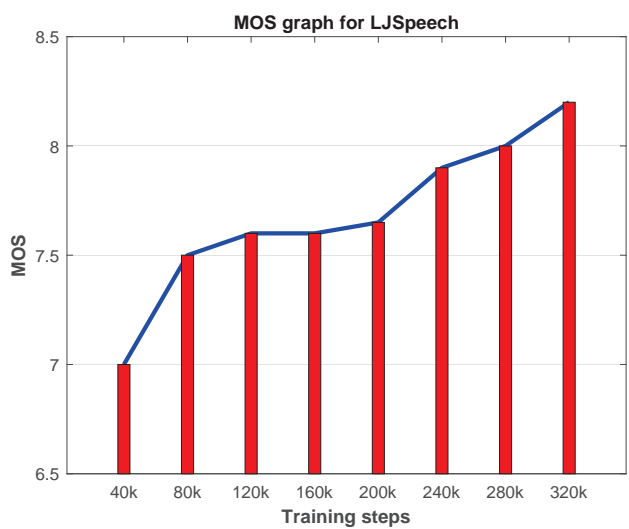}	
			\caption{Quality analysis of synthesized speech for LJSpeech dataset}
			\label{qa_ljspeech}
		\end{center}
	\end{figure}

\begin{figure}[]
		\begin{center}
			\includegraphics[width=0.67\textwidth]{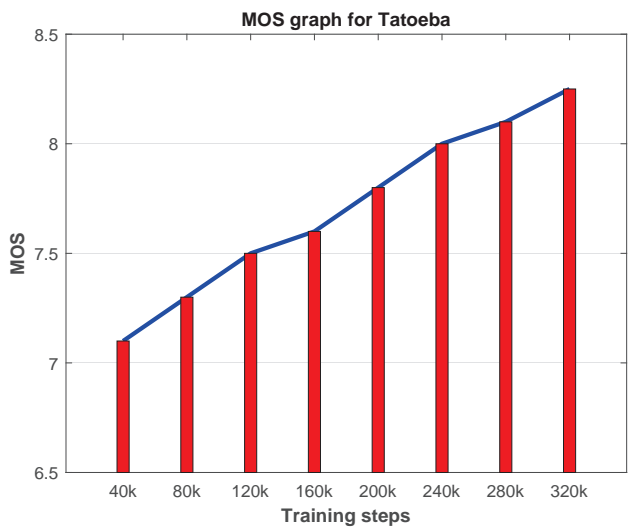}
			\caption{Quality analysis of synthesized speech for Tatoeba dataset}	
			\label{qa_tope}
		\end{center}
	\end{figure}

\begin{figure}[]
	\begin{center}
		\includegraphics[width=0.67\textwidth]{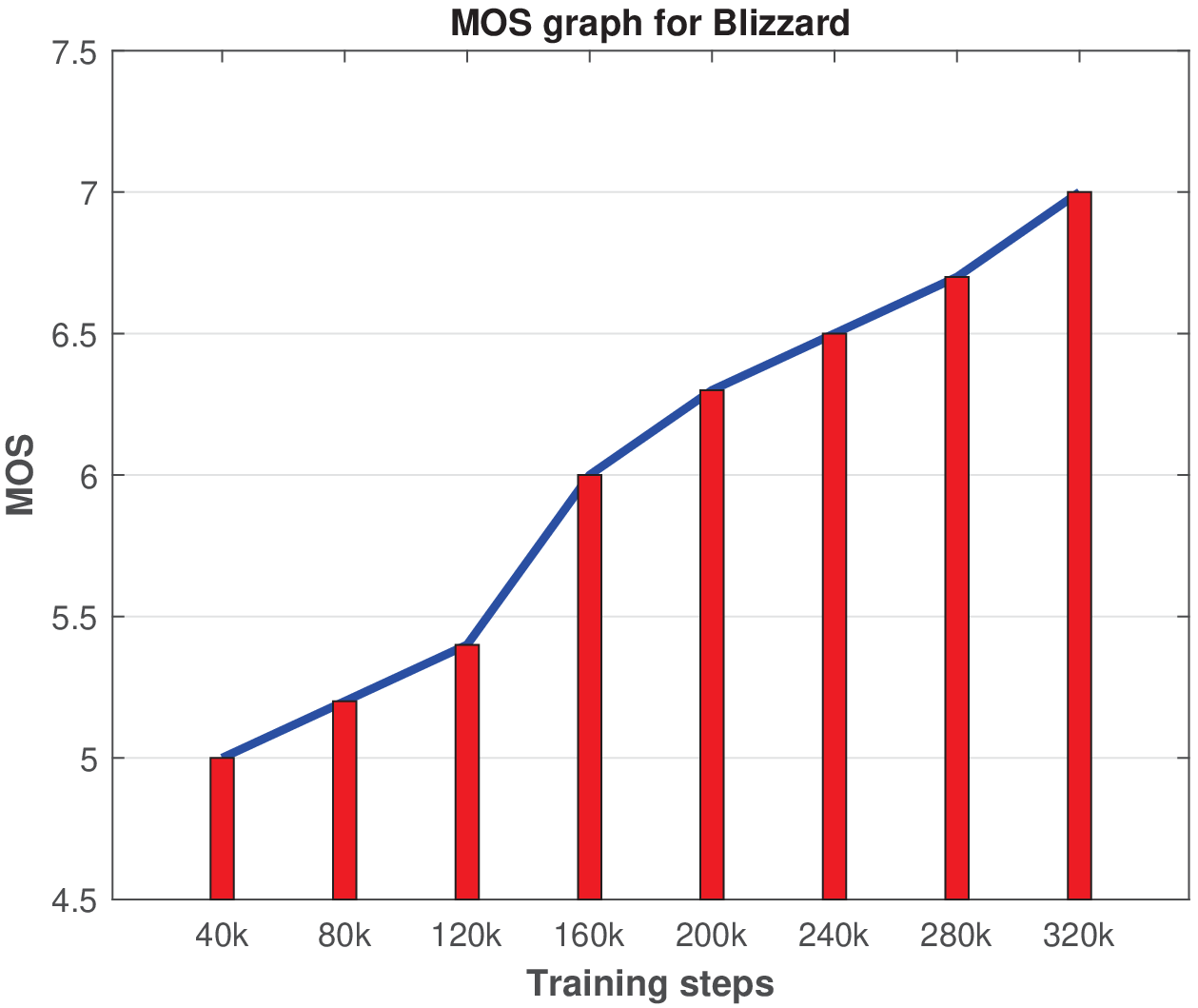}	
		\caption{Quality analysis of synthesized speech for Blizzard dataset}
		\label{qa_blizzard}
	\end{center}
	\end{figure}

	Quality analysis results have been shown in Table~\ref{tab:ca} and~\ref{tab:moe}. It was observed that, after 250k steps of model training, the output speech included prosody features. That made the voice more feasible for real-time speech synthesis. Speech quality also depends on the corpus used for training. Model is trained up to 400K steps for all datasets and results have been generated. The quality of the synthesized speech is expected to be clear to understand and non-robotic in nature. The more human-like the voice is, have a higher value of MOS and results easy to understand. X axis shows duration of model trained and Y axis represents MOS of speech taken by 10 evaluators. The output speech has been evaluated at regular training step intervals for different data sets. The speech quality was observed to depend mainly on the number of iterations used in the algorithm. The graph shows that in the initial stage learning rate is very high but after 250k training steps learning rate is very slow. The observed results in context of the quality of the output speech are visually illustrated in Fig.~\ref{qa_ljspeech},~\ref{qa_tope} and ~\ref{qa_blizzard}.\vspace{.3cm} \\	
	\textbf{Accuracy of MOS Determination}\vspace{.2cm}\\
	To further evaluate the quality of the proposed methodology's results, 100 text samples with known MOS scores are considered. Corresponding speech is synthesized for them using GLA, FGLA and GAN based vocoders. MOS of the synthesized speech have been evaluated. If the variation in the MOS of the synthesized speech i.e. $MOS_s$ and the known MOS i.e. $MOS_g$ is less than the error margin $e$, then the sample is assumed to be accurately determined. The error margin is taken as 0.45 which is 5\% of the MOS of natural human voice~\cite{MOSGT}. The calculations are done as per Eq.~\ref{eqc} and the results for the considered samples are summaried in Table~\ref{tab:MOSCompare}. \vspace{-.2cm}

	{\fontsize{9}{8.75}\selectfont
	\begin{eqnarray}
	\label{eqc}
	\begin{split}		
	\centering
			&
			\Big\{
			|MOS_g – MOS_s| < \textit{e} => Accurately\ determined.
			\Big\}
	&\\Where:\\
	&\begin{cases}
	\textit{e}:error\ margin.\\
	MOS_g:ground\ truth\ MOS.\\
	MOS_s:MOS\ of\ synthesized\ speech.\\
	\end{cases}		
	\end{split}
	\end{eqnarray}
	\vspace{-.2cm}

	\linespread{1.25}	
	\begin{table}[H]
	\begin{center}
		\renewcommand{\arraystretch}{1.25} 
		\caption{Accuracy of MOS determination }\vspace{-.1cm}
		\label{tab:MOSCompare}
		\begin{tabular}{|l|c|c|c|c|}
			\hline
			\textbf{Vocoder}     & \multicolumn{1}{l|}{\textbf{Avg. MOS}} & \multicolumn{1}{l|}{\textbf{Min. MOS}} & \multicolumn{1}{l|}{\textbf{Max. MOS}} & \multicolumn{1}{l|}{\textbf{Accuracy}} \\ \hline
			\textbf{GLA}         & 7.5                                    & 5.2                                    & 7.6                                    & 72\%                                   \\ \hline
			\textbf{FGLA}        & 7.8                                    & 5.4                                    & 8.2                                    & 81\%                                   \\ \hline
			\textbf{GAN Vocoder} & 7.6                                    & 4.8                                    & 8.0                                    & 79\%                                   \\ \hline
		\end{tabular}
	\end{center}
	\end{table}\vspace{-.2cm}

	The accuracy for FGLA came out to be much better than GLA and comparable to GAN based neural vocoder. It should be noted that GAN vocoder has been evaluated on CPU, in-line with the goal of optimizing the waveform processing from linear spectrograms for real-time devices with limited processing capabilities.
}

	\subsubsection{Synthesis Delay Analysis}\label{res:time} 
	`Synthesis Delay' is the time required for the output speech to start getting produced by the TTS system. The same has been computed and compared for all the use-case sentences described in Table~\ref{tab:tc} and datasets mentioned in Section~\ref{dsets}. The observations have been drawn for GLA with 30 iterations, FGLA with 60 iterations and GAN based neural vocoder~\cite{oyamada2018generative}. Every sentence is synthesised 10 times and then average synthesis delay has been calculated. Same computational configuration has been maintained on the testing machines while doing so. As our aim is to optimize the waveform processing keeping CPU based low memory devices for real-time usage, the waveform reconstruction has been carried out on CPU for GLA, FGLA and the GAN based vocoder. 
	
\begin{figure}[]
		\begin{center}
			\includegraphics[width=0.95\textwidth]{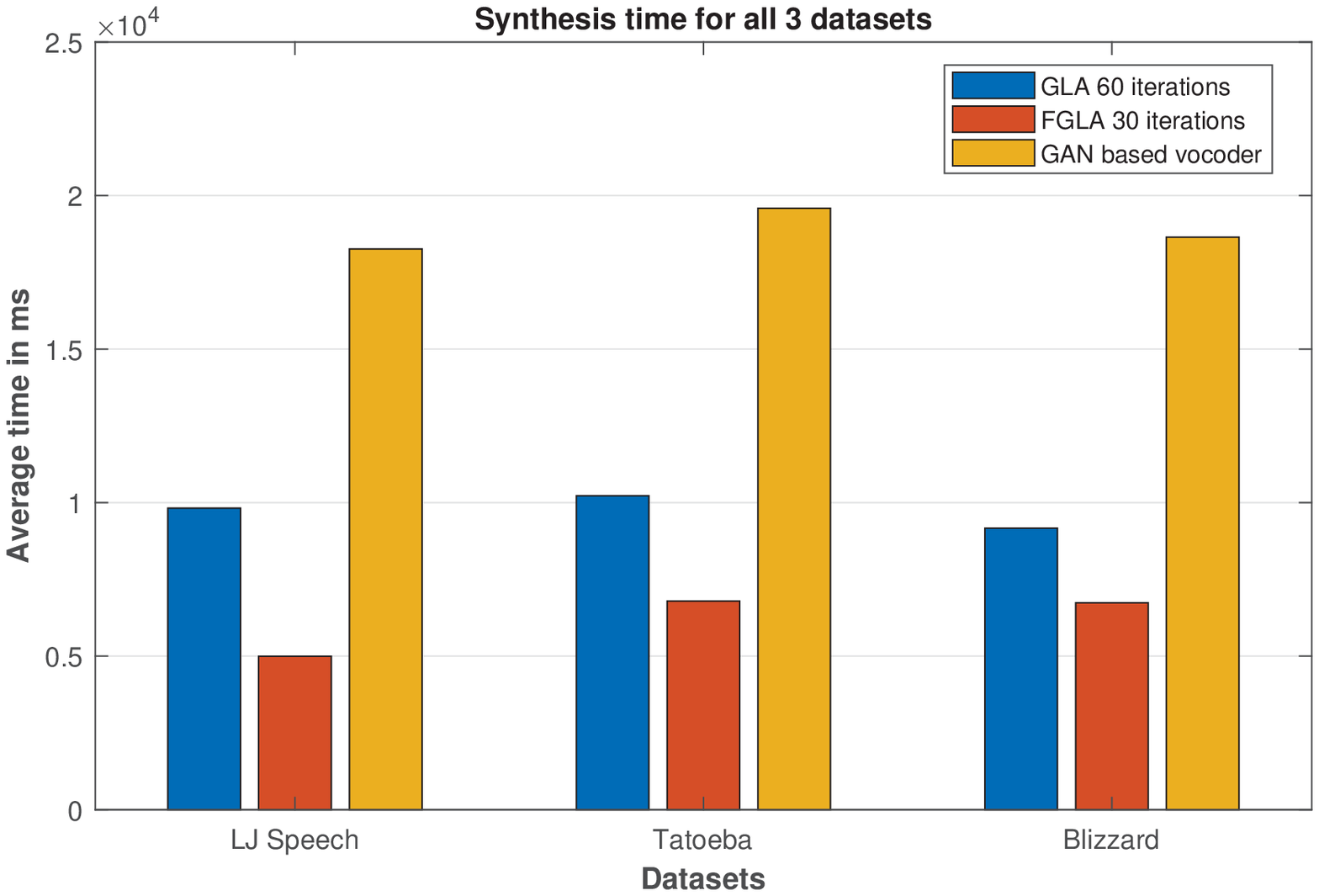}	
			\caption{Synthesis Delay for the use-cases formulated in Table~\ref{tab:tc}}
			\label{synth_time}
		\end{center}
	\end{figure}

	The observed synthesis time for the aforementioned cases have been depicted numerically in Table~\ref{tab:usecase} and visually in Fig.~\ref{synth_time}. FGLA came out to be 49.12\%, 33.57\% and 26.52\% faster than GLA in terms of Synthesis Delay for LJSpeech, Tatoeba and Blizzard datasets. The overall reduction in the Synthesis Delay has been observed to be 36.58\%. While the average speech synthesis time for GAN based vocoder on CPU came out to be 3.65, 2.88 and 2.76 times more than FGLA. It should also be noted that FGLA produced better quality speech with lesser training iterations as compared to GLA. With the proposed waveform generation strategy, LJ Speech dataset has shown more reduction in the synthesis delay than other datasets. In context of MOS scores, Tatoeba dataset showed faster progress while Blizzard dataset showed lower values as compared to LJSpeech.\\

	\linespread{1.25}	
	\begin{table}[]
		\caption{Synthesis time (ms) for various use-cases and datasets}
		\label{tab:usecase}
		\begin{center}
			\resizebox{.65
				\textwidth}{!}{%
				\begin{adjustbox}{angle=90}
					\renewcommand{\arraystretch}{2}
				\begin{tabular}{|l|p{6.2cm}|C{1cm}|C{1cm}|C{1.15cm}|C{1cm}|C{1cm}|C{1.15cm}|C{1cm}|C{1cm}|C{1.15cm}|}
					\hline
					\multicolumn{1}{|c|}{\multirow{2}{*}{\textbf{S.N.}}}& \multicolumn{1}{c|}{\multirow{2}{*}{\textbf{Use-case Sentences}}} & \multicolumn{3}{c|}{\textbf{LJSpeech}} & \multicolumn{3}{c|}{\textbf{Tatoeba}} & \multicolumn{3}{c|}{\textbf{Blizzard}} \\ \cline{3-11} 
					& & \textbf{GLA (60)} & \textbf{FGLA (30)} & \textbf{GAN vocoder} & \textbf{GLA (60)} & \textbf{FGLA (30)} & \textbf{GAN vocoder} & \textbf{GLA (60)} & \textbf{FGLA (30)} & \textbf{GAN vocoder} \\ \hline
						1. & He said to him, “Is not your name Ahmed?" & 10026 & 5173 &18900 & 9684 & 6456 &17950 & 9800 & 6725 & 16850\\ \hline
						2. & All of a sudden, there was a loud screaming, Please help me! & 9698 & 4940 &18190 & 10416 & 6880 &19950 & 8997 & 6726 &18870 \\ \hline
						3. & I think I lost my wallet! I can’t find it anywhere! Oh, I could just kick myself! & 9702 & 4920 & 17000 & 10180 & 6799 & 20010 & 8957 & 6702 & 19010 \\ \hline
						4. & ``Sunshine on my shoulders makes me happy, sunshine in my eyes can make me cry." & 9849 & 4959 &18700 & 10550 & 7010 & 19880 & 8940 & 6757 &18550 \\ \hline
						5. & As the stranger entered the town, he was met by a police, man who asked, “Are you a traveler?" “So it would appear", He replied carelessly. & 9828 & 4992 & 18560 & 10300 & 6821 &20150 & 9129 & 6763 &19950   \\ \hline
									\end{tabular}%
								\end{adjustbox}
						}\end{center}
					\end{table}

	\subsubsection{Convergence Plots}\label{convplots}		
\begin{figure}[]
		\centering
		\begin{subfigure}[b]{0.5\textwidth}            
			\includegraphics[width=\textwidth]{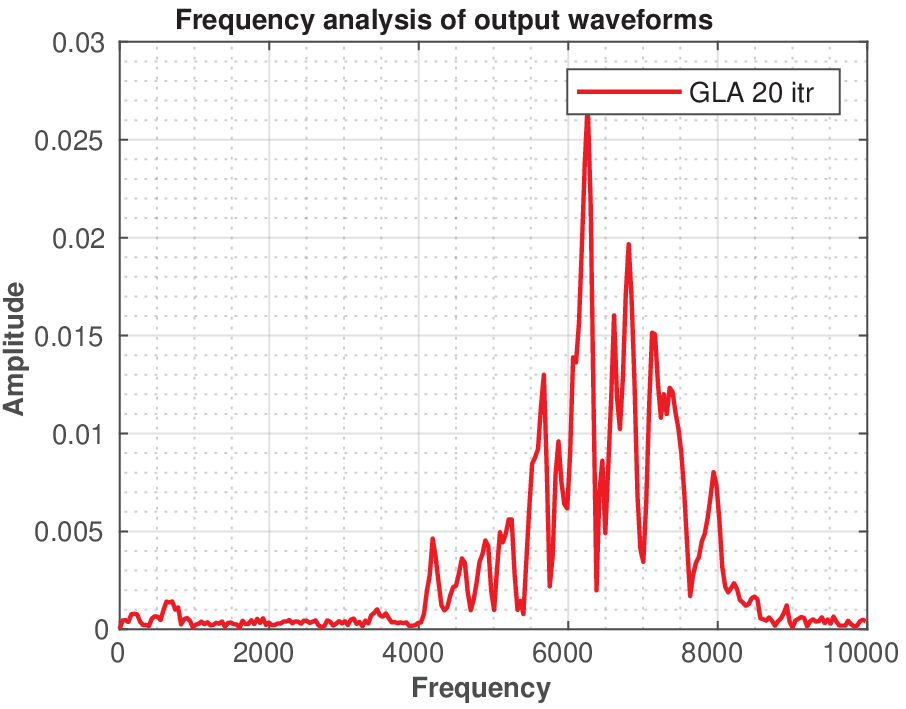}
			\caption{GLA 20 Iterations}
			\label{fig:plotg1}
		\end{subfigure}% 
		\begin{subfigure}[b]{0.5\textwidth}
			\centering
			\includegraphics[width=\textwidth]{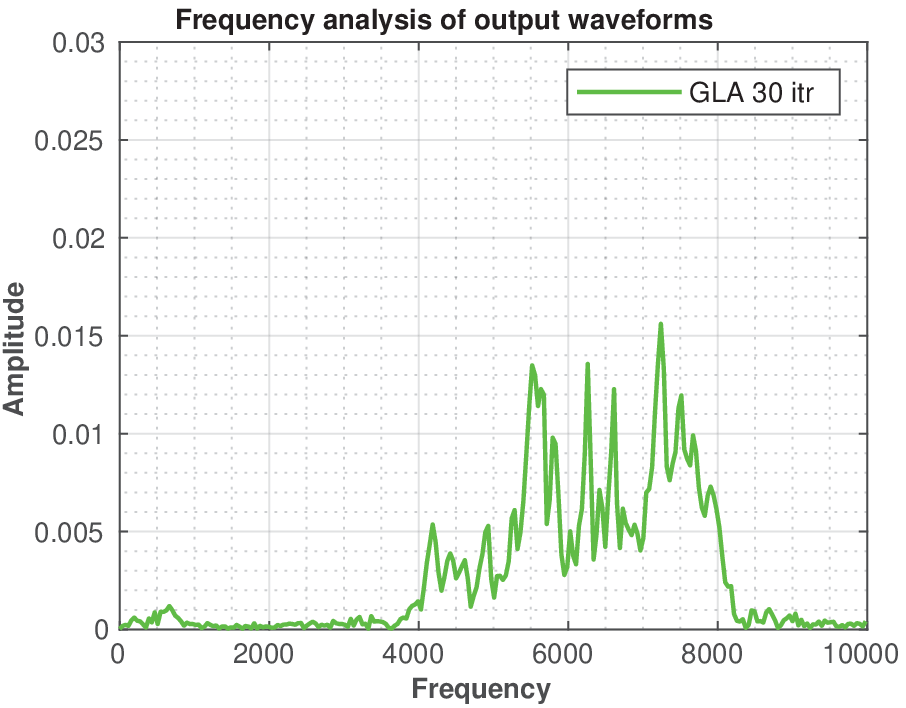}
			\caption{GLA 30 Iterations}
			\label{fig:plotg2}
		\end{subfigure}\vspace{.2in}
		\begin{subfigure}[c]{0.5\textwidth}
			\centering
			\includegraphics[width=\textwidth]{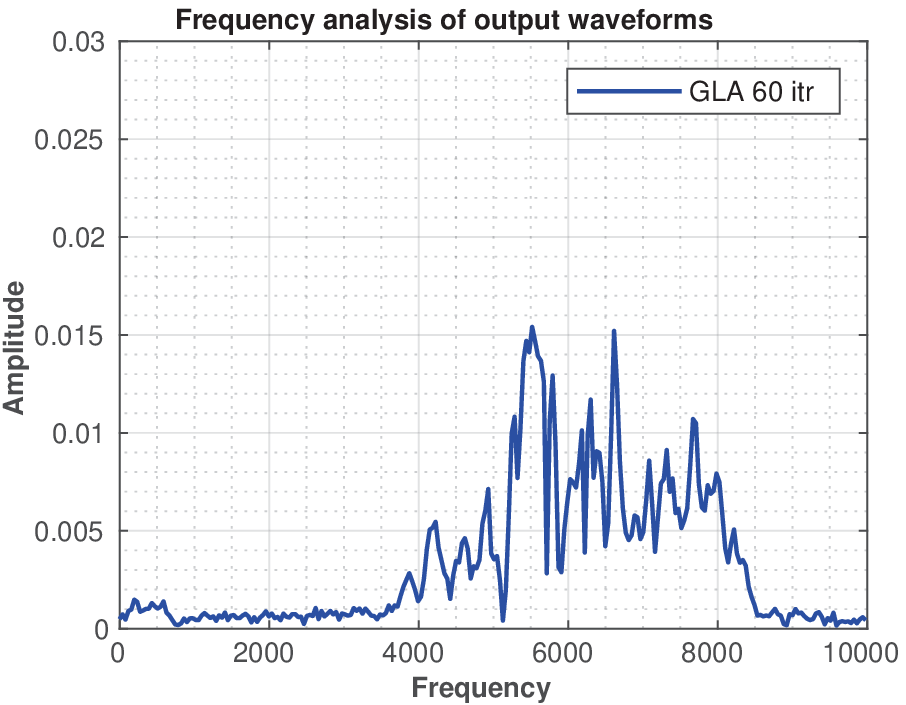}
			\caption{GLA 60 Iterations}
			\label{fig:plotg3}
		\end{subfigure}% 
		\begin{subfigure}[d]{0.5\textwidth}
			\centering
			\includegraphics[width=\textwidth]{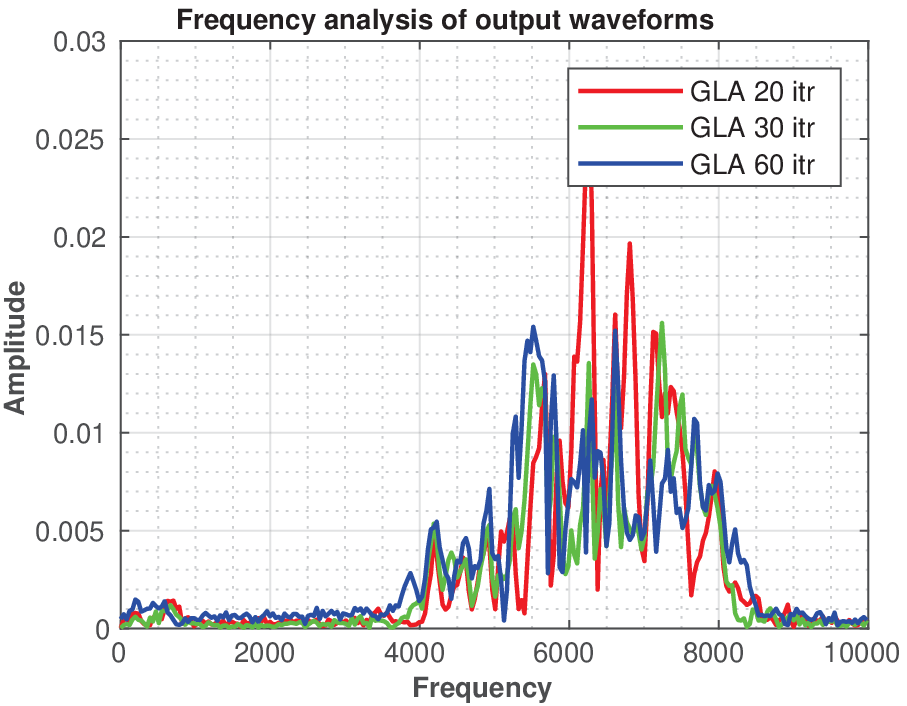}
			\caption{GLA 20, 30 \& 60 Iterations}
			\label{fig:plotg4}
		\end{subfigure}
		\caption{GLA Convergence Plots}\label{fig:gla_convplot}
	\end{figure}

\begin{figure}[]
	\centering
	\begin{subfigure}[b]{0.5\textwidth}            
		\includegraphics[width=\textwidth]{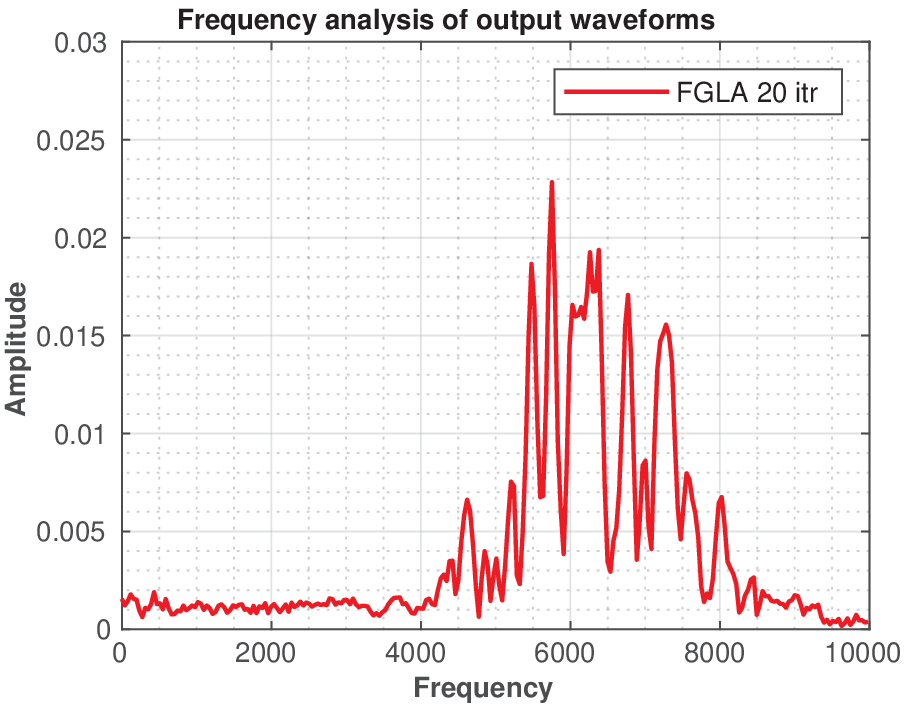}
		\caption{FGLA 20 Iterations}
		\label{fig:plotf1}
	\end{subfigure}% 
	\begin{subfigure}[b]{0.5\textwidth}
		\centering
		\includegraphics[width=\textwidth]{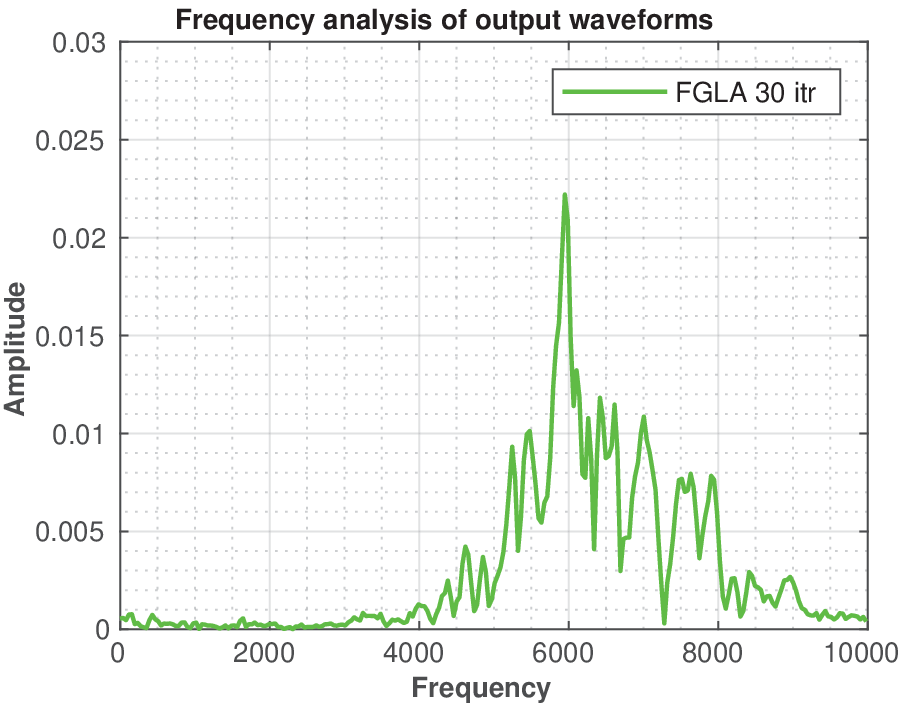}
		\caption{FGLA 30 Iterations}
		\label{fig:plotf2}
	\end{subfigure}\vspace{.2in}
	\begin{subfigure}[c]{0.5\textwidth}
		\centering
		\includegraphics[width=\textwidth]{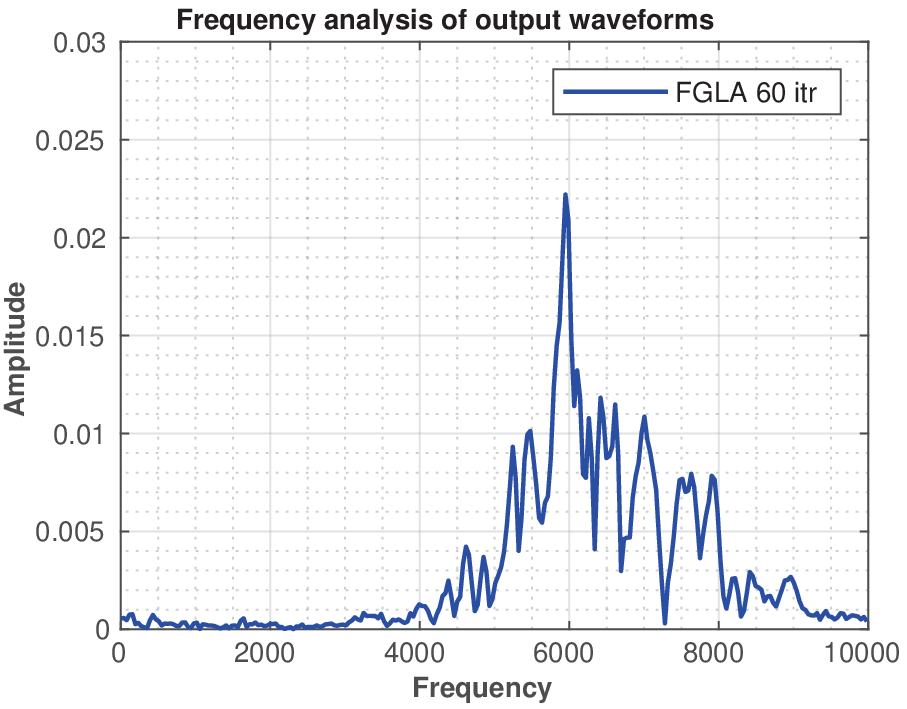}
		\caption{FGLA 60 Iterations}
		\label{fig:plotf3}
	\end{subfigure}% 
	\begin{subfigure}[d]{0.5\textwidth}
		\centering
		\includegraphics[width=\textwidth]{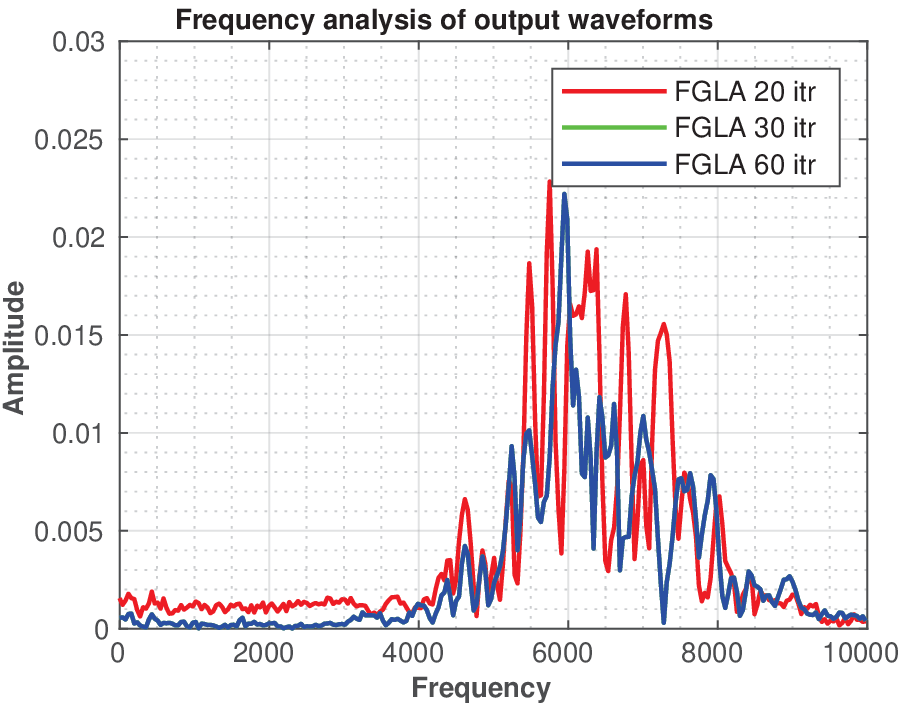}
		\caption{FGLA 20, 30 \& 60 Iterations}
		\label{fig:plotf4}
	\end{subfigure}
	\caption{FGLA Convergence Plots}\label{fig:fgla_convplot}
	\end{figure}

	As discussed in Section~\ref{itr}, appropriate number of iterations for FGLA were determined as 30. Here, we have verified that choice by observing the convergence in the plots of the resulting waveforms. The comparative analysis in the frequency domain is easier to observe. That's why, fourier transformations of the waveforms are considered. The transforms of the speech produced from FLGA with 30 iterations and FLGA with 60 iterations are plotted and compared. It is found that the two plots overlap with each other. Similarly, fourier transforms of waveforms produced by GLA 60 iterations and GLA 30 iterations are plotted and compared. It is found that the plots do not overlap. This means that the waveform produced by FLGA 30 iterations and FGLA 60 iterations are same, thus a good quality of speech is produced using FLGA 30 itself, whereas, using GLA requires 60 iterations give a better quality of speech that GLA 60 iterations. The convergence plots for the fourth sentence from Table.~\ref{tab:tc} are shown in Fig.~\ref{fig:gla_convplot} and Fig.~\ref{fig:fgla_convplot} for GLA and FGLA respectively. The fourth sentence captures sufficient variations in terms of sentence length, special characters, punctuation marks, etc. The plots for the rest of the sentences are included in the supplementary material.
	
	%%%%%%%%%%%%%%%%%%%%%%%%%%%%% Conclusion & Future Work %%%%%%%%%%%%%%%%%%%%%%%%%%%%%%%	
	\section{Conclusion}\label{con}
	In this paper, a FGLA based method is proposed to optimize the waveform generation process to reduce speech synthesis delay. The final speech, i.e., waveform, is reconstructed from intermediate spectrogram. GLA has mostly been used, especially when phase information about the waveform is missing. But GLA is slow, which causes delay in the speech synthesis. A faster alternative of GLA, i.e., FGLA has been used in the proposed method that resulted in 36.58\% reduction in speech synthesis time. In the presented work, experiments were performed to optimize waveform generation from linear spectrogram in single-speaker TTS systems. The proposed approach is compared against GLA and GAN based neural vocoder in terms of speech quality and synthesis delay. The quality of the synthesized speech has been checked using MOS based evaluation. The quality is observed to be retained in spite of the reduction in the synthesis time. The number of training steps and iterations were determined by experimental observation. This choice was verified through the convergence of fourier transform plots of the resultant waveforms. 
	
	In future, we will work to optimize the waveform processing for the TTS systems trained with multi-speaker datasets using mel-spectrograms as intermediate representation. We will also explore more neural vocoders for speech synthesis in real-time applications. It is planned to work on the challenges involved with them such as reducing the model size and reducing the computational requirements especially for synthesizing small sentences.

	%%%%%%%%%%%%%%%%%%%%%%%%%%%%%%%%% References %%%%%%%%%%%%%%%%%%%%%%%%%%%%%%%%%
	\bibliographystyle{unsrt}
	\bibliography{Manuscript}
	
\end{document}